\documentclass[aps,prd,twocolumn,superscriptaddress,showpacs]{revtex4}
\usepackage{epsfig,epsf}
\usepackage{amsmath}
\usepackage{amsthm}
\usepackage{amsfonts}
\usepackage{amssymb}
\usepackage{dsfont}
\usepackage{multirow}
\usepackage{appendix}
\usepackage{slashed}
\usepackage[active]{srcltx}
\usepackage{psfrag}

\begin{document}

\title{ Resonance $X(5568)$ as an exotic axial-vector state }
\date{\today}
\author{S.~S.~Agaev}
\affiliation{Institute for Physical Problems, Baku State University, Az--1148 Baku,
Azerbaijan}
\author{K.~Azizi}
\affiliation{Department of Physics, Do\v{g}u\c{s} University, Acibadem-Kadik\"{o}y, 34722
Istanbul, Turkey}
\author{B.~Barsbay}
\affiliation{Department of Physics, Kocaeli University, 41380 Izmit, Turkey}
\author{H.~Sundu}
\affiliation{Department of Physics, Kocaeli University, 41380 Izmit, Turkey}

\begin{abstract}
The mass and meson-current coupling constant of the resonance $X(5568)$, as
well as the width of the decay $X(5568)\to B_s^{\ast}\pi$ are calculated by
modeling the exotic $X(5568)$ resonance as a diquark-antidiquark state $%
X_b=[su][bd]$ with quantum numbers $J^{P}=1^{+}$. The calculations are made
employing QCD two-point sum rule method, where the quark, gluon and mixed
vacuum condensates up to dimension eight are taken into account. The sum
rule approach on the light-cone in its soft-meson approximation is used to
explore the vertex $X_bB_{s}^{\ast}\pi$ and extract the strong coupling $%
g_{X_bB_{s}^{\ast}\pi}$, which is a necessary ingredient to find the width
of the $X_b \to B_s^{\ast}\pi^{+}$ decay process. The obtained predictions
are compared with the experimental data of the D0 Collaboration, and results
of other theoretical works.
\end{abstract}

\pacs{12.39.Mk, 14.40.Rt, 14.40.Pq}
\maketitle

\section{Introduction}

Investigation of the "exotic" hadrons, which cannot be described as usual $%
q\bar q$ and $qqq$ structures, and are composed of more than three valence
quarks is now one of the intriguing and developing branches of high energy
physics. Existence of such particles, their possible properties, and
experiments suitable for their observation were among interests of
physicists during last three decades. Starting from discovery of the theory
of strong interactions, i.e. the Quantum Chromodynamics (QCD), a qualitative
analysis of the exotic states were replaced by the quantitative calculations
of their parameters in the strong context of the quantum field theory. Some
of parameters of the exotic states were computed in eighties using namely
the nonperturbative tools of QCD in Refs.\ \cite{FirstTheor1,FirstTheor2}.
But theoretical studies of the exotic particles then were not supported by a
reliable experimental information, which slowed the growth of the field.

Situation changed, when the various collaborations, such as Belle, BaBar,
BESIII, LHCb, CDF, and D0 started to supply the hadron physics by the
valuable experimental data on the quantum numbers, masses, and decay widths
of new heavy resonances, some of which were interpreted as four-quark (i.e.,
as tetraquark) exotic states \cite{FirstExp}. Experimental studies,
naturally, boosted suggestion of new theoretical models to explain an
internal structure of the observed resonances, as well as led to invention
of new or adapting existing nonperturbative approaches to meet problems of
the new emerging field of hadron physics. It is worth noting that during
last years a considerable success was achieved both in the experimental and
theoretical studies of the exotic states (see, the reviews \cite%
{Jaffe:2004ph,Swanson:2006st,Klempt:2007cp,Godfrey:2008nc,Voloshin:2007dx,Nielsen:2010, Faccini:2012pj,Esposito:2014rxa,Chen:2016}
and references therein).

Among the family of the tetraquarks the narrow resonance $X(5568)$ has a
unique position. The evidence for existence of these resonance was announced
recently by the D0 Collaboration \cite{D0:2016mwd}, which was founded on the
$p\bar{p}$ collision data at $\sqrt{s}=1.96\ \mathrm{TeV}$. It was observed
in the channel $X(5568) \to B_{s}^{0} \pi^{\pm}$ through chain of decays $%
B_{s}^{0} \to J/\psi \phi$, $J/\psi \to \mu^{+} \mu^{-}$, and $\phi \to
K^{+}K^{-}$. It is not difficult to conclude that the state $X(5568)$
consists of valence $b,\, s, \, u$ and $d$ quarks, and is presumably the
first observed particle built of four different quarks. The measured mass of
this state equals to $m_{X}=5567.8 \pm 2.9 \mathrm{(stat)}^{+0.9}_{-1.9}
\mathrm{(syst)}\, \mathrm{MeV}$, and its decay width is $\Gamma=21.9 \pm 6.4
\mathrm{(stat)}^{+5.0}_{-2.5} \mathrm{(syst)}\, \mathrm{MeV}$. The D0
assigned to this particle the quantum numbers $J^{P}=0^{+}$. But if the
resonance $X(5568)$ decays as $X(5568) \to B_{s}^{\ast}\pi^{\pm} \to
B_{s}^{0} \gamma \pi^{\pm}$ with unseen soft $\gamma$, then the difference $%
m(B_{s}^{\ast})-m(B_{s}^{0})$ should be added to the mass of the resonance $%
X(5568)$, whereas its decay width remains unchanged \cite{D0:2016mwd}. In
this case the resonance has the quantum numbers $J^{P}=1^{+}$. Stated
differently, in accordance with the results of the D0 Collaboration the
resonance $X(5568)$ may be treated as the particle with $J^{P}=0^{+}$ or $%
J^{P}=1^{+}$.

The decay channel $B_{s}^{0}\pi^{\pm}$ was investigated by the LHCb
Collaboration utilizing the $pp$ collision data at energies $7\, \mathrm{TeV}
$ and $8\, \mathrm{TeV}$ collected at CERN \cite{LHCb:2016}. The aim was to
confirm the existence of the $X(5568)$ state and measure its spectroscopic
parameters. But the LHCb Collaboration could not fix the resonance structure
in the $B_s^{0}\pi^{\pm} $ invariant mass distribution at the energies less
than $6000\ \, \mathrm{MeV}$. Very similar conclusions were drawn by the CMS
collaboration \cite{CMS:2016}, in which a mass range up to $5900 \ \,
\mathrm{MeV}$ was searched for a possible structure, setting an upper limit.
Nevertheless, the D0 Collaboration from analysis of the semileptonic decays of
$B_{s}^0$ meson recently confirmed in Ref.\ \cite{D0} the observation of the $X(5568)$
resonance. As is seen, experimental situation with the existence of the exotic state $%
X(5568)$, supposedly built of four different quark flavors, remains intriguing and unclear.

Namely these conditions make the theoretical studies of the $X(5568)$ state
even more urgent than just after first announcement of its observation.
Suggestions about inner organization of this particle as the bound state of
a diquark and antidiquark, or as a meson molecule compound were already made
in Ref.\ \cite{D0:2016mwd}. Theoretical works, appeared afterwards to
determine the mass and decay width of the $X(5568)$ state followed mainly
these suggestions. Thus, in Ref.\ \cite{Agaev:2016mjb} it was considered as
the diquark-antidiquark structure $X_b=[su][\bar{b}\bar{d}]$ with the
quantum numbers $0^{++}$, where its mass $m_{X_b}$ and meson-current
coupling (hereafter, the meson coupling) $f_{X_b}$ were calculated. In the
framework of the diquark-antidiquark model the mass and other parameters of $%
X(5568)$ were also explored in Refs.\ \cite%
{Wang:2016tsi,Chen:2016mqt,Zanetti:2016wjn,Wang:2016mee}. The values for $%
m_{X_b}$ found in these works agree with each other, and are consistent with
the experimental data of the D0 Collaboration.

The width of the $X_b \to B_{s}^{0}\pi^{+}$ decay channel was calculated in
Ref.\ \cite{Agaev:2016ijz} employing for $X_b$ the same structure and
interpolating current as in Ref.\ \cite{Agaev:2016mjb}. To this end, authors
applied QCD light-cone sum rule method and soft-meson approximation adjusted
for studying of the tetraquark states in Ref.\ \cite{Agaev:2016dev}, where
relevant explanations and technical details can be found. The result for $%
\Gamma(X_{b}^{+} \to B_{s}^{0}\pi^{+})$ derived in Ref.\ \cite{Agaev:2016ijz}
describes correctly the experimental data. The width of the decay channels $%
X^{\pm}(5568) \to B_{s}\pi^{\pm}$ was also analyzed in Refs.\ \cite%
{Dias:2016dme,Wang:2016wkj} employing the three-point QCD sum rule approach.
In these works authors found a nice agreement between the theoretical
predictions for $\Gamma(X^{\pm} \to B_{s}^{0}\pi^{\pm})$ and data, as well.

As we have mentioned above, the $X(5568)$ state can also be considered as a
molecule composed of $B$ and $\overline{K}$ mesons, which was realized in
Refs.\ \cite{Xiao:2016mho,Agaev:2016urs,Chen:2016ypj}. But in this scenario,
in accordance with Ref.\ \cite{Agaev:2016urs}, the mass of such molecule and
width of the decay $X^{\pm}(5568) \to B_{s}\pi^{\pm}$ exceed the
experimental data of D0, and predictions of the diquark-antidiquark model.
These facts were interpreted in favor of the diquark-antidiquark
organization of the $X(5568)$ state.

The experimental data of the D0 and LHCb collaborations generated an
appearance of interesting theoretical works, where the structure and
spectroscopic parameters, production mechanisms of the $X(5568)$ state were
considered. The details of used methods, necessary explanations, and
conclusions made there concerning a nature of the resonance $X(5568)$ can be
found in the original papers \cite%
{Agaev:2016srl,He:2016yhd,Jin:2016cpv,Stancu:2016sfd,Burns:2016gvy,Tang:2016pcf, Guo:2016nhb,Lu:2016zhe, Esposito:2016itg,Albaladejo:2016eps,Ali:2016gdg,Kang:2016,Lang:2016}.

In the present work we calculate the mass, meson coupling of the $X_{b}=[su][%
\bar{b}\bar{d}]$ diquark-antidiquark state treating it as an axial-vector
particle with the quantum numbers $J^{P}=1^{+}$. To this end, we apply QCD
two-point sum rule approach and include into our analysis the quark, gluon
and mixed condensates up to eight dimensions. We are also going to determine
the width of the decay $X_{b}\rightarrow B_{s}^{\ast }\pi ^{+}$ using the
soft-meson version of QCD light-cone sum rule method. Performed
investigation should allow us to answer a question is the $X(5568)$
resonance a state with $J^{P}=0^{+}$ or $J^{P}=1^{+}$.

This work is organized in the following form. Section \ref{sec:Mass} ie
devoted to the sum rule calculations of the mass and meson coupling of the
axial vector $X_b$ state. Here we derive also the light-cone sum rule
expression for the strong coupling $g_{X_bB_{s}^{\ast}\pi}$ necessary to
compute the decay width of the process $X_b \to B_{s}^{\ast}\pi^{+}$. In
Sect.\ \ref{sec:Num} we conduct numerical calculations and extract values of
the mass, meson coupling and decay width under consideration. Here we
compare our results with the experimental data and predictions of other
theoretical works. Section \ref{sec:Conc}contains our brief conclusions on
the nature of the $X_b$ state based on the present studies. The explicit
expressions for the spectral density used in the two-point sum rules
are moved to Appendix \ref{sec:App}.


\section{The QCD sum rules for the parameters of the axial-vector $X_b$ state%
}

\label{sec:Mass}
In this section we derive QCD sum rules required for both the mass and decay
width calculations. To this end, we calculate the two-point spectral density
necessary for the mass and meson coupling computations. To
extract the width of the decay channel $X_b \to B_{s}^{\ast}\pi^{+}$,  we find
the strong coupling $g_{X_bB_{s}^{\ast}\pi}$ using the
spectral densities corresponding  to different Lorentz structures in the
corresponding correlation function.

\subsection{Sum rules for the mass and meson coupling}

In order to find QCD two-point sum rules for calculation of the mass and
meson coupling of the $X_{b}$ state we consider the correlation function
given below as
\begin{equation}
\Pi _{\mu \nu }(q)=i\int d^{4}xe^{iq\cdot x}\langle 0|\mathcal{T}\{J_{\mu
}(x)J_{\nu }^{\dag }(0)\}|0\rangle ,  \label{eq:CorrF1}
\end{equation}%
where $J_{\mu }(x)$ is the interpolating current with the quantum numbers of
the $X_b$ state. We consider $X_{b}$ as a particle with the quantum numbers $%
J^{P}=1^{+}$. Then in the diquark-antidiquark model one of the acceptable
interpolating currents $J_{\mu }(x)$ is defined by the expression \cite%
{Chen:2016mqt}
\begin{eqnarray}
&&J_{\mu }(x)=s_{a}^{T}(x)C\gamma _{5}u_{b}(x)\left\{ \overline{b}%
_{a}(x)\gamma _{\mu }C\overline{d}_{b}^{T}(x)\right.  \notag \\
&&\left. -\overline{b}_{b}(x)\gamma _{\mu}C\overline{d}_{a}^{T}(x)\right\} ,
\label{eq:CDiq}
\end{eqnarray}%
where $a$ and $b$ are color indices and $C$ is the charge conjugation matrix.

In order to derive QCD sum rule expression we first have to calculate the
correlation function in terms of the physical parameters of the $X_b$ state.
Saturating the correlation function with a complete set of the $X_b$ state
and performing integral over $x$ in Eq.\ (\ref{eq:CorrF1}), we get
\begin{equation}
\Pi _{\mu \nu }^{\mathrm{Phys}}(q)=\frac{\langle 0|J_{\mu }|X_{b}(q)\rangle
\langle X_{b}(q)|J_{\nu }^{\dagger }|0\rangle }{m_{X_{b}}^{2}-q^{2}}+...
\end{equation}%
with $m_{X_{b}}$ being the mass of the $X_{b}$ state. Here the dots indicate
contributions to the correlation function arising from the higher resonances
and continuum states. We define the meson coupling $f_{X_{b}}$ using the
matrix element
\begin{equation}
\langle 0|J_{\mu }|X_{b}(q)\rangle =f_{X_{b}}m_{X_{b}}\varepsilon _{\mu },
\label{eq:Res}
\end{equation}%
where $\varepsilon _{\mu }$ is the polarization vector of the $X_{b}$ state.
Then in terms of $m_{X_{b}}$ and $f_{X_{b}}$, the correlation function can
be written in the form
\begin{equation}
\Pi _{\mu \nu }^{\mathrm{Phys}}(q)=\frac{m_{X_{b}}^{2}f_{X_{b}}^{2}}{%
m_{X_{b}}^{2}-q^{2}}\left( -g_{\mu \nu }+\frac{q_{\mu }q_{\nu }}{%
m_{X_{b}}^{2}}\right) +\ldots  \label{eq:CorM}
\end{equation}%
Equation (\ref{eq:CorM}) has been derived within the single pole assumption,
which in the case of a multiquark state requires additional justification.
The reason is that for such systems in the physical side of the sum rule one
has to take into account also two-hadron reducible contributions, which for
the tetraquarks however are negligible \cite{Matheus:2009}. Therefore, we
can apply the Borel transformation directly to Eq.\ (\ref{eq:CorM}), which
yields
\begin{equation}
\mathcal{B}_{q^{2}}\Pi _{\mu \nu }^{\mathrm{Phys}%
}(q)=m_{X_{b}}^{2}f_{X_{b}}^{2}e^{-m_{X_{b}}^{2}/M^{2}}\left( -g_{\mu \nu }+%
\frac{q_{\mu }q_{\nu }}{m_{X_{b}}^{2}}\right) +\ldots  \label{eq:CorBor}
\end{equation}

From the QCD side the same function must be calculated employing of the
quark-gluon degrees of freedom. To this end, we contract the heavy and light
quark fields and find for the correlation function $\Pi _{\mu \nu }^{\mathrm{%
QCD}}(q)$ the following expression:
\begin{eqnarray}
&&\Pi _{\mu \nu }^{\mathrm{QCD}}(q)=i\int d^{4}xe^{iqx}\left\{ \mathrm{Tr}%
\left[ \gamma _{5}\widetilde{S}_{s}^{aa^{\prime }}(x)\right. \right.  \notag
\\
&&\left. \times \gamma _{5}S_{u}^{bb^{\prime }}(x)\right] \mathrm{Tr}\left[
\gamma _{\mu }\widetilde{S}_{d}^{a^{\prime }b}(-x)\gamma _{\nu
}S_{b}^{b^{\prime }a}(-x)\right]  \notag \\
&&-\mathrm{Tr}\left[ \gamma _{\mu }\widetilde{S}_{d}^{b^{\prime
}b}(-x)\gamma _{\nu }S_{b}^{a^{\prime }a}(-x)\right] \mathrm{Tr}\left[
\gamma _{5}\widetilde{S}_{s}^{aa^{\prime }}(x)\right.  \notag \\
&&\times \left. \gamma _{5}S_{u}^{bb^{\prime }}(x)\right] +\mathrm{Tr}\left[
\gamma _{5}\widetilde{S}_{s}^{aa^{\prime }}(x)\gamma _{5}S_{u}^{bb^{\prime
}}(x)\right]  \notag \\
&&\times \mathrm{Tr}\left[ \gamma _{\mu }\widetilde{S}_{d}^{b^{\prime
}a}(-x)\gamma _{\nu }S_{b}^{a^{\prime }b}(-x)\right] -\mathrm{Tr}\left[
\gamma _{\mu }\widetilde{S}_{d}^{a^{\prime }a}(-x)\right.  \notag \\
&&\left. \times \left. \gamma _{\nu }S_{b}^{b^{\prime }b}(-x)\right] \mathrm{%
Tr}\left[ \gamma _{5}\widetilde{S}_{s}^{aa^{\prime }}(x)\gamma
_{5}S_{u}^{bb^{\prime }}(x)\right] \right\} .  \label{eq:CorrF2}
\end{eqnarray}%
In Eq.\ (\ref{eq:CorrF2}) $S_{s(u,d)}^{ab}(x)$ and $S_{b}^{ab}(x)$ are the
light $s$, $u$, $d$ quarks and $b$-quark propagators, respectively. Here we
also use the notation
\begin{equation*}
\widetilde{S}_{s(d)}(x)=CS_{s(d)}^{T}(x)C.
\end{equation*}%
We work with the light quark propagator $S_{q}^{ab}(x)$ defined in the form%
\begin{eqnarray}
&&S_{q}^{ab}(x)=i\delta _{ab}\frac{\slashed x}{2\pi ^{2}x^{4}}-\delta _{ab}%
\frac{m_{q}}{4\pi ^{2}x^{2}}-\delta _{ab}\frac{\langle \overline{q}q\rangle
}{12}  \notag \\
&&+i\delta _{ab}\frac{\slashed xm_{q}\langle \overline{q}q\rangle }{48}%
-\delta _{ab}\frac{x^{2}}{192}\langle \overline{q}g_{s}\sigma Gq\rangle
+i\delta _{ab}\frac{x^{2}\slashed xm_{q}}{1152}  \notag \\
&&\times \langle \overline{q}g_{s}\sigma Gq\rangle-i\frac{g_sG_{ab}^{\alpha
\beta }}{32\pi ^{2}x^{2}}\left[ \slashed x{\sigma _{\alpha \beta }+\sigma
_{\alpha \beta }}\slashed x\right]  \notag \\
&&-i\delta _{ab}\frac{x^{2}\slashed xg_{s}^{2}\langle \overline{q}q\rangle
^{2}}{7776}-\delta _{ab}\frac{x^{4}\langle \overline{q}q\rangle \langle
g_{s}^{2}G^2\rangle }{27648}+\ldots  \label{eq:qprop}
\end{eqnarray}%
Let us emphasize that in calculations we set the light quark masses $m_u$
and $m_d$ equal to zero, preserving at the same time dependence of the
propagator $S_{s}^{ab}(x)$ on the $m_s$. For the $b$-quark propagator $%
S_{b}^{ab}(x)$ we employ the formula given in Ref.\ \cite{Reinders:1984sr}
\begin{eqnarray}
&&S_{b}^{ab}(x)=i\int \frac{d^{4}k}{(2\pi )^{4}}e^{-ikx} \Bigg \{ \frac{%
\delta _{ab}\left( {\slashed k}+m_{b}\right) }{k^{2}-m_{b}^{2}}  \notag \\
&&-\frac{g_{s}G_{ab}^{\alpha \beta }}{4}\frac{\sigma _{\alpha \beta }\left( {%
\slashed k}+m_{b}\right) +\left( {\slashed k}+m_{b}\right) \sigma _{\alpha
\beta }}{(k^{2}-m_{b}^{2})^{2}}  \notag \\
&&+\frac{g_{s}^{2}G^{2}}{12}\delta _{ab}m_{b}\frac{k^{2}+m_{b}{\slashed k}}{%
(k^{2}-m_{b}^{2})^{4}}+\frac{g_{s}^{3}G^{3}}{48}\delta _{ab}\frac{\left( {%
\slashed k}+m_{b}\right) }{(k^{2}-m_{b}^{2})^{6}}  \notag \\
&& \times \left[ {\slashed k}\left( k^{2}-3m_{b}^{2}\right) +2m_{b}\left(
2k^{2}-m_{b}^{2}\right) \right] \left( {\slashed k}+m_{b}\right) +\ldots %
\Bigg \}.  \notag \\
&& {}  \label{eq:Qprop}
\end{eqnarray}%
In Eqs.\ (\ref{eq:qprop}) and (\ref{eq:Qprop}) we use the notations
\begin{eqnarray}
&&G_{ab}^{\alpha \beta } = G_{A}^{\alpha \beta
}t_{ab}^{A},\,\,~~G^{2}=G_{\alpha \beta }^{A}G_{\alpha \beta }^{A},  \notag
\\
&&G^{3} =\,\,f^{ABC}G_{\mu \nu }^{A}G_{\nu \delta }^{B}G_{\delta \mu }^{C},
\end{eqnarray}%
where $a,\,b=1,2,3$ and $A,B,C=1,\,2\,\ldots 8$ are the color indices. Here $%
t^{A}=\lambda ^{A}/2$, and $\lambda ^{A}$ are the Gell-Mann matrices. In the
nonperturbative terms the gluon field strength tensor $G_{\alpha \beta
}^{A}\equiv G_{\alpha \beta }^{A}(0)$ is fixed at $x=0.$

The correlation function $\Pi _{\mu \nu }^{\mathrm{QCD}}(q)$ can be
decomposed over the Lorentz structures $\sim g_{\mu \nu }$ and $\sim q_{\mu
}q_{\nu }$. The QCD sum rule expressions can be obtained after fixing the
same Lorentz structures in the both $\Pi _{\mu \nu }^{\mathrm{Phys}}(q)$ and
$\Pi _{\mu \nu }^{\mathrm{QCD}}(q)$. We choose the term $\sim g_{\mu \nu }$,
which receives a contribution from only spin-$1$ states, whereas the
invariant amplitude corresponding to the structure $\sim q_{\mu} q_{\nu}$
forms due to contributions of both spin-$0$ and spin-$1$ states.

The chosen invariant amplitude $\Pi ^{\mathrm{QCD}}(q^{2})$ can be written
down as the dispersion integral
\begin{equation}
\Pi ^{\mathrm{QCD}}(q^{2})=\int_{(m_{b}+m_{s})^{2}}^{\infty }\frac{\rho ^{%
\mathrm{QCD}}(s)}{s-q^{2}}ds+...,
\end{equation}%
where $\rho ^{\mathrm{QCD}}(s)$ is the two-point spectral density. Now
utilizing the Borel transformation to $\Pi ^{\mathrm{QCD}}(q^{2})$ ,
equating the obtained expression with the relevant part of the function $%
\mathcal{B}_{q^{2}}\Pi _{\mu \nu }^{\mathrm{Phys}}(q)$, and subtracting the
continuum contribution we, as a result, get the required sum rule. Then the
mass of the $X_{b}$ state can be evaluated from the sum rule
\begin{equation}
m_{X_{b}}^{2}=\frac{\int_{(m_{b}+m_{s})^{2}}^{s_{0}}dss\rho ^{\mathrm{QCD}%
}(s)e^{-s/M^{2}}}{\int_{(m_{b}+m_{s})^{2}}^{s_{0}}ds\rho (s)e^{-s/M^{2}}}.
\label{eq:srmass}
\end{equation}%
To extract the meson coupling $f_{X_{b}}$ we can employ the sum rule formula
\begin{equation}
m_{X_{b}}^{2}f_{X_{b}}^{2}e^{-m_{X_{b}}^{2}/M^{2}}=%
\int_{(m_{b}+m_{s})^{2}}^{s_{0}}ds\rho ^{\mathrm{QCD}}(s)e^{-s/M^{2}}.
\label{eq:srcoupling}
\end{equation}

The key component of these expressions is the two-point
spectral density $\rho^{\mathrm{QCD}}(s)$.  Because the methods for its
derivation were presented in the literature (see, Ref.\ \cite{Agaev:2016dev}%
), here we omit technical details of calculations moving the final expressions of the
spectral density and its components to Appendix \ref{sec:App}.

\subsection{Strong coupling $g_{X_bB_{s}^{\ast}}\protect\pi$ and width of
the decay $X_b \to B_s^{\ast}\protect\pi^{+}$}

In this subsection we outline the soft-meson approximation of QCD light-cone
sum rule method used here to explore the strong vertex $X_bB_{s}^{\ast}\pi$
and find the sum rule expression for the coupling $g_{X_bB_{s}^{\ast}\pi}$.
The latter will be used to calculate the width of the $X_b \to
B_s^{\ast}\pi^{+}$ decay process.

To this end, we start from the correlation function
\begin{equation}
\Pi_{\mu \nu}(p,q)=i\int d^{4}xe^{ipx}\langle \pi (q)|\mathcal{T}%
\{J_{\mu}^{B_{s}^{\ast}}(x)J_{\nu}^{\dag }(0)\}|0\rangle ,  \label{eq:CorrF3}
\end{equation}
where the interpolating current for the $B_{s}^{\ast}$ meson has the form
\begin{equation}
J_{\mu}^{B_{s}^{\ast}}(x)=\overline{b}_{l}(x)\gamma _{\mu}s_{l}(x),
\label{eq:Bcur}
\end{equation}
whereas $J_{\nu }(x)$ is defined by Eq.\ (\ref{eq:CDiq}). Here $p$, $q$ and $%
p^{\prime }=p+q$ are the momenta of the mesons $B_{s}^{\ast}$ and $\pi $,
and the $X_b$ state, respectively.

To derive the sum rule for the strong coupling $g_{X_{b}B_{s}^{\ast }\pi }$,
we follow the standard prescriptions of the QCD sum rule approach and
calculate $\Pi _{\mu \nu }(p,q)$ in terms of the physical parameters of the
involving particles. Then we obtain
\begin{eqnarray}
\Pi _{\mu \nu }^{\mathrm{Phys}}(p,q) &=&\frac{\langle 0|J_{\mu
}^{B_{s}^{\ast }}|B_{s}^{\ast }\left( p\right) \rangle }{p^{2}-m_{B_{s}^{%
\ast }}^{2}}\langle B_{s}^{\ast }\left( p\right) \pi (q)|X_{b}(p^{\prime
})\rangle  \notag \\
&&\times \frac{\langle X_{b}(p^{\prime })|J_{\nu }^{\dag}|0\rangle }{%
p^{\prime 2}-m_{X_{b}}^{2}}+\ldots .  \label{eq:CorrF4}
\end{eqnarray}%
where the dots denote, as usual, contributions of the higher resonances and
continuum states. The expression of the function $\Pi _{\mu \nu }^{\mathrm{%
Phys}}(p,q)$ can be further simplified and expressed in terms of the
particle parameters if introduce the new matrix elements
\begin{eqnarray}
&&\langle 0|J_{\mu }^{B_{s}^{\ast }}|B_{s}^{\ast }\left( p\right) \rangle
=f_{B_{s}^{\ast }}m_{B_{s}^{\ast }}\varepsilon _{\mu },  \notag \\
&&\langle B_{s}^{\ast }\left( p\right) \pi (q)|X_{b}(p^{\prime })\rangle =
\left[ (p\cdot p^{\prime })(\varepsilon ^{\ast }\cdot \varepsilon ^{\prime
})\right.  \notag \\
&&\left. -(p\cdot \varepsilon ^{\prime })(p^{\prime }\cdot \varepsilon
^{\ast })\right] g_{X_{b}B_{s}^{\ast }\pi },  \label{eq:Mel}
\end{eqnarray}%
where $f_{B_{s}^{\ast }},$ $m_{B_{s}^{\ast }},$ $\varepsilon _{\mu }$ are
the decay constant, mass and polarization vector of the $B_{s}^{\ast }\left(
p\right) $ meson, and $\varepsilon _{\nu }^{\prime }$ is the polarization
vector of the $X_{b}$ state.

Using these matrix elements, as well as one given by Eq.\ (\ref{eq:Res}),
one can rewrite the correlation function as
\begin{eqnarray}
&&\Pi _{\mu \nu }^{\mathrm{Phys}}(p,q)=\frac{f_{B_{s}^{\ast
}}f_{X_{b}}m_{X_{b}}m_{B_{s}^{\ast }}}{\left( p^{\prime
2}-m_{X_{b}}^{2}\right) \left( p^{2}-m_{B_{s}^{\ast }}^{2}\right) }%
g_{X_{b}B_{s}^{\ast }\pi }  \notag \\
&&\times \left( \frac{m_{X_{b}}^{2}+m_{B_{s}^{\ast }}^{2}}{2}g_{\mu \nu
}-p_{\mu }^{\prime }p_{\nu }\right) +\ldots   \label{eq:CorrF5}
\end{eqnarray}%
It evidently contains two Lorentz structures $g_{\mu \nu }$ and $p_{\mu
}^{\prime }p_{\nu }$. We are going to use both of them to derive the sum
rules for the coupling $g_{X_{b}B_{s}^{\ast }\pi }$ and compare our results
with each other in order to estimate a sensitivity of the obtained
predictions to the chosen structures. Below we consider explicitly the
correlation function corresponding to $g_{\mu \nu }$, and provide only final
expression for the spectral density $\widetilde{\rho }_{c}^{\mathrm{QCD}}(s)$
derived using the terms $\sim p_{\mu }^{\prime }p_{\nu }$.

In the soft-meson limit, i.e. in the limit $q=0$, the Borel transformation
of the invariant amplitude corresponding to $g_{\mu \nu }$ has the form
[Ref.\ \cite{Agaev:2016dev}],
\begin{eqnarray}
&&\Pi ^{\mathrm{Phys}}(M^{2})=f_{B_{s}^{\ast
}}f_{X_{b}}m_{X_{b}}m_{B_{s}^{\ast }}g_{X_{b}B_{s}^{\ast }\pi }m^{2}  \notag
\\
&&\times \frac{1}{M^{2}}e^{-m^{2}/M^{2}}+\cdots ,  \label{eq:BT}
\end{eqnarray}%
where $m^{2}=(m_{X_{b}}^{2}+m_{B_{s}^{\ast }}^{2})/2$.

It is known, that to obtain the sum rule expression for the strong coupling
of the three-hadron vertex, in the soft-meson approximation one has to apply
the one-variable Borel transformation instead of the two-variable Borel
transformation accepted in the standard light-cone sum rule procedures \cite%
{Agaev:2016dev}. As a results, the hadronic part of the sum rule becomes
more complicated and contains contributions of transitions from exited
states to the ground level, shown in Eq.\ (\ref{eq:BT}) as the dots, which
in the soft-meson approximation are unsuppressed even after Borel
transformation. In order to remove these terms from the hadronic side one
has to use some technical tools. One of such methods was suggested in Ref.\
\cite{Ioffe:1983ju}, which implies acting by the operator
\begin{equation}
\left( 1-M^{2}\frac{d}{dM^{2}}\right) M^{2}e^{m^{2}/M^{2}}  \label{eq:softop}
\end{equation}%
to both the hadronic and QCD sides of the sum rule. It was successfully used
in Ref.\ \cite{Agaev:2016dev}, and is adopted in the present work, as well.

But before that we have to calculate the correlation function Eq.\ (\ref%
{eq:CorrF3}) in terms of the quark propagators and find QCD side of the sum
rule. Contractions of $s$ and $b$-quark fields in Eq.\ (\ref{eq:CorrF3})
give
\begin{eqnarray}
&&\Pi _{\mu \nu }^{\mathrm{QCD}}(p,q)=i\int d^{4}xe^{ipx}\left\{ \left[
\gamma _{5}\widetilde{S}_{s}^{ia}(x){}\gamma _{\mu }\right. \right.  \notag
\\
&&\left. \times \widetilde{S}_{b}^{bi}(-x){}\gamma _{\nu }\right] _{\alpha
\beta }\langle \pi (q)|\overline{u}_{\alpha }^{b}d_{\beta }^{a}|0\rangle
\notag \\
&&\left. -\left[ \gamma _{5}\widetilde{S}_{s}^{ia}(x){}\gamma _{\mu }%
\widetilde{S}_{b}^{ai}(-x){}\gamma _{\nu }\right] _{\alpha \beta }\langle
\pi (q)|\overline{u}_{\alpha }^{b}d_{\beta }^{b}|0\rangle \right\} ,
\label{eq:CorrF6}
\end{eqnarray}%
where $\alpha $ and $\beta $ are the spinor indices, the quark fields are
defined at $x=0$.

In the soft-meson approximation the QCD side of the sum rule is considerably
simpler than one in the standard approach. Indeed, in the standard
light-cone sum rule the non-local matrix elements of the hadrons (in the
case under analysis, of the pion) are expressed in terms of its various
distribution amplitudes, whereas in the soft-meson limit we get only few
local matrix elements.

The following stages of calculations include applying of the expansion
\begin{equation}
\overline{u}_{\alpha }^{b}d_{\beta }^{a}\rightarrow \frac{1}{4}\Gamma
_{\beta \alpha }^{j}\left( \overline{u}^{b}\Gamma ^{j}d^{a}\right) ,
\label{eq:MatEx}
\end{equation}%
where $\Gamma ^{j}$ is the full set of Dirac matrixes, performing the color
summation and inserting into the quark matrix elements the gluon field
strength tensor $G$. These operations lead to the correlation function $\Pi
_{\mu \nu }^{\mathrm{QCD}}(p,q)$, which depends on the two and
three-particle matrix elements of the pion: Let us note that in the present
work we neglect terms $\sim G^{2}$ and $\sim G^{3}$. The final step in this
way is computation of the imaginary part of the obtained correlation
function to extract the desired spectral density. Because these manipulation
are described in a clear form in Ref.\ \cite{Agaev:2016dev}, we refrain from
providing further details and write down the final expression for $\rho
_{c}^{\mathrm{QCD}}(s)$ obtained in this work as the sum its the
perturbative and nonperturbative parts
\begin{equation}
\rho _{\mathrm{c}}^{\mathrm{QCD}}(s)=\rho _{\mathrm{c}}^{\mathrm{pert.}%
}(s)+\rho _{\mathrm{c}}^{\mathrm{n.-pert.}}(s),  \label{eq:SD}
\end{equation}%
where
\begin{equation}
\rho _{\mathrm{c}}^{\mathrm{pert.}}(s)=\frac{f_{\pi }\mu _{\pi }}{48\pi
^{2}s^{2}}(m_{b}^{2}-s)(m_{b}^{4}+m_{b}^{2}s-6m_{b}m_{s}s-2s^{2})
\label{eq:SD1}
\end{equation}%
and
\begin{eqnarray}
&&\rho _{\mathrm{c}}^{\mathrm{n.-pert.}}(s)=\frac{f_{\pi }\mu _{\pi }}{12}%
\langle \overline{s}s\rangle \left[ sm_{s}\delta
^{^{(1)}}(s-m_{b}^{2})\right.  \notag \\
&&\left. +(m_{s}-2m_{b})\delta (s-m_{b}^{2})\right] +\frac{f_{\pi }\mu _{\pi
}}{72}\langle \overline{s}g_{s}\sigma Gs\rangle  \notag \\
&&\times \left\{ 3(2m_{b}-m_{s})\delta
^{^{(1)}}(s-m_{b}^{2})+s(3m_{b}-5m_{s})\right.  \notag \\
&&\left. \times \delta ^{(2)}(s-m_{b}^{2})-s^2m_{s}\delta
^{(3)}(s-m_{b}^{2})\right\} .  \label{eq:SD2}
\end{eqnarray}%
The contributions $\sim \delta ^{(n)}(s-m_{b}^{2})=(d/ds)^{n}\delta
(s-m_{b}^{2})$ arise from the imaginary part of the pole terms.

Calculations of the spectral density $\widetilde{\rho }_{\mathrm{c}}^{%
\mathrm{QCD}}(s)$ performed employing the terms, which in the soft limit
transform to ones $\sim p'_{\mu }p_{\nu }$, yield:
\begin{equation}
\widetilde{\rho }_{\mathrm{c}}^{\mathrm{pert.}}(s)=\frac{f_{\pi }\mu _{\pi }%
}{24\pi ^{2}s^{3}}(s^{3}-3sm_{b}^{4}+2m_{b}^{6})  \label{eq:SD3}
\end{equation}%
and
\begin{eqnarray}
&&\widetilde{\rho }_{\mathrm{c}}^{\mathrm{n.-pert.}}(s)=-\frac{f_{\pi }\mu
_{\pi }}{6}m_{s}\langle \overline{s}s\rangle \delta ^{^{(1)}}(s-m_{b}^{2})
\notag \\
&&+\frac{f_{\pi }\mu _{\pi }}{36}m_{s}\langle \overline{s}g_{s}\sigma
Gs\rangle \left[ 4\delta ^{(2)}(s-m_{b}^{2})\right.   \notag \\
&&\left. +s\delta ^{(3)}(s-m_{b}^{2})\right] .  \label{eq:SD4}
\end{eqnarray}

As is seen, the spectral densities depend on the parameters $f_{\pi } $ and $%
\mu _{\pi }$ of the pion through the local matrix element
\begin{equation}
\langle 0|\overline{d}i\gamma _{5}u|\pi (q)\rangle =f_{\pi }\mu _{\pi },
\label{eq:MatE2}
\end{equation}%
where
\begin{equation}
\mu _{\pi }=\frac{m_{\pi }^{2}}{m_{u}+m_{d}}=-\frac{2\langle \overline{q}%
q\rangle }{f_{\pi }^{2}}.  \label{eq:PionEl}
\end{equation}

Performing the continuum subtraction in the standard manner and applying the
operator Eq.\ (\ref{eq:softop}), we get the sum rule to evaluate the strong
coupling
\begin{eqnarray}
&&g_{X_{b}B_{s}^{\ast }\pi }=\frac{1}{f_{B_{s}^{\ast
}}f_{X_{b}}m_{X_{b}}m_{B_{s}^{\ast }}m^{2}}\left( 1-M^{2}\frac{d}{dM^{2}}%
\right) M^{2}  \notag \\
&&\times \int_{(m_{b}+m_{s})^{2}}^{s_{0}}dse^{(m^{2}-s)/M^{2}}\rho _{c}^{%
\mathrm{QCD}}(s).  \label{eq:SRules1}
\end{eqnarray}%
The similar computations for the structure  $p'_{\mu }p_{\nu }$ give:
\begin{eqnarray}
&&\widetilde{g}_{X_{b}B_{s}^{\ast }\pi }=\frac{1}{f_{B_{s}^{\ast
}}f_{X_{b}}m_{X_{b}}m_{B_{s}^{\ast }}}\left( 1-M^{2}\frac{d}{dM^{2}}\right)
M^{2}  \notag \\
&&\times \int_{(m_{b}+m_{s})^{2}}^{s_{0}}dse^{(m^{2}-s)/M^{2}}\widetilde{%
\rho }_{c}^{\mathrm{QCD}}(s).
\label{eq:SRules2}
\end{eqnarray}

The width of the decay $X_{b}\rightarrow B_{s}^{\ast }\pi ^{+}$ can be borrowed from Ref.\ \cite%
{Agaev:2016dev}, and is identical
for  both the strong couplings $g_{X_{b}B_{s}^{\ast }\pi }$ and $\widetilde{g}%
_{X_{b}B_{s}^{\ast }\pi }$
\begin{eqnarray}
&&\Gamma \left( X_{b}\rightarrow B_{s}^{\ast }\pi ^{+}\right) =\frac{%
g_{X_{b}B_{s}^{\ast }\pi }^{2}m_{B_{s}^{\ast }}^{2}}{24\pi }\lambda \left(
m_{X_{b}},\ m_{B_{s}^{\ast }},m_{\pi }\right)   \notag \\
&&\times \left[ 3+\frac{2\lambda ^{2}\left( m_{X_{b}},\ m_{B_{s}^{\ast
}},m_{\pi }\right) }{m_{B_{s}^{\ast }}^{2}}\right] ,  \label{eq:DW}
\end{eqnarray}%
where
\begin{equation*}
\lambda (a,\ b,\ c)=\frac{\sqrt{a^{4}+b^{4}+c^{4}-2\left(
a^{2}b^{2}+a^{2}c^{2}+b^{2}c^{2}\right) }}{2a}.
\end{equation*}%
The final expressions (\ref{eq:SRules1}), (\ref{eq:SRules2}) and (\ref{eq:DW})
will be used for numerical analysis of the decay channel $X_{b}\rightarrow B_{s}^{\ast }\pi
^{+}$.


\section{Numerical results}

\label{sec:Num}
The QCD sum rules contain numerous parameters, i.e. quark, gluon and mixed
condensates, masses of the $b$ and $s$ quarks, and $B_{s}^{\ast}$ meson's
mass and decay constant $f_{B_s^{\ast}}$. Values of these parameters are
written down in Table \ref{tab:Param}. Let us note, that
we use the well known values for the quark-gluon condensates and fix them within 
usual procedures. The gluon condensate $\langle g_{s}^3G^3\rangle$, which is
not commonly employed in the sum rule calculations, is borrowed from Ref.\ \cite%
{Narison:2015nxh}. Further we choose the mass of the $b$ quark in the $\overline{MS}$
scheme at the scale $\mu=m_b$, whereas for the decay constant $%
f_{B_s^{\ast}} $ invoke the result presented in Ref.\ \cite{Dae}. Other
parameters are taken from Ref.\ \cite{Agashe:2014kda}.

\begin{table}[tbp]
\begin{tabular}{|c|c|}
\hline\hline
Parameters & Values \\ \hline\hline
$m_{B_s^{\ast}}$ & $(5415.4_{-2.1}^{+2.4}~\mathrm{MeV}$ \\
$m_{B_s^{0}}$ & $(5366.77\pm0.24)~\mathrm{MeV}$ \\
$f_{B_s^{\ast}}$ & $(225\pm 9)~\mathrm{MeV}$ \\
$m_{\pi}$ & $139.57 ~\mathrm{MeV}$ \\
$f_{\pi}$ & $0.131~\mathrm{GeV}$ \\
$m_{b}$ & $(4.18\pm0.03)~\mathrm{GeV}$ \\
$m_{s} $ & $(95 \pm 5)~\mathrm{MeV} $ \\
$\langle \bar{q}q \rangle $ & $(-0.24\pm 0.01)^3$ $\mathrm{GeV}^3$ \\
$\langle \bar{s}s \rangle $ & $0.8\ \langle \bar{q}q \rangle$ \\
$m_{0}^2 $ & $(0.8\pm0.1)$ $\mathrm{GeV}^2$ \\
$\langle \overline{s}g_s\sigma Gs\rangle$ & $m_{0}^2\langle \bar{s}s \rangle$
\\
$\langle\frac{\alpha_sG^2}{\pi}\rangle $ & $(0.012\pm0.004)$ $~\mathrm{GeV}%
^4 $ \\
$\langle g_{s}^3G^3\rangle $ & $(0.57\pm0.29)$ $~\mathrm{GeV}^6 $ \\
\hline\hline
\end{tabular}%
\caption{Input parameters.}
\label{tab:Param}
\end{table}

The QCD sum rule expressions depend also on the continuum threshold $s_{0}$
and Borel parameter $M^{2}$. To extract values of the quantities under
consideration we have to choose such regions for these parameters, where
dependence of the physical quantities under consideration on them is
minimal. In practice, however we may only to reduce effect connected with
our choices of the windows for $s_{0}$ and $M^{2}$. 

The QCD sum rule method, as we know,  suffers from the theoretical uncertainties,
which are its unavoidable property. The main sources of ambiguities in extracting 
the physical quantities under question are the continuum threshold $s_0$ and Borel 
parameter $M^2$. But procedures to extract these parameters are well defined  
in the context of the sum rule method itself, where obtained results depend
on an accuracy of the calculations.

Here some comments are necessary concerning the theoretical accuracy achieved
in the present work in deriving the spectral  density $\rho^{\mathrm{QCD}}(s)$.
In fact, there are some features of $\rho^{\mathrm{QCD}}(s)$,
which differ it from the existing calculations. First, all its higher dimensional
components $\rho_{k}(s)$, including $\rho_{7}(s)$ one, are nonzero and contribute to
$\rho^{\mathrm{QCD}}(s)$. Secondly, apart from the quark, gluon and mixed condensates
$\langle \overline{q}q\rangle$, $\langle \alpha_{s}G^{2}/\pi \rangle$ and
$\langle \overline{q}g_s\sigma Gq\rangle$, the spectral density $\rho^{\mathrm{QCD}}(s)$
encompasses effects of the terms $\sim \langle \alpha _{s}G^{2}/{\pi }\rangle ^{2}$ and
$\sim \langle g_{s}^{3}G^{3}\rangle$ appearing due to more detailed formulas for the
quark propagators accepted in the present work.

\begin{figure}[tbp]
\centerline{
\begin{picture}(210,170)(0,0)
\put(0,0){\epsfxsize8.2cm\epsffile{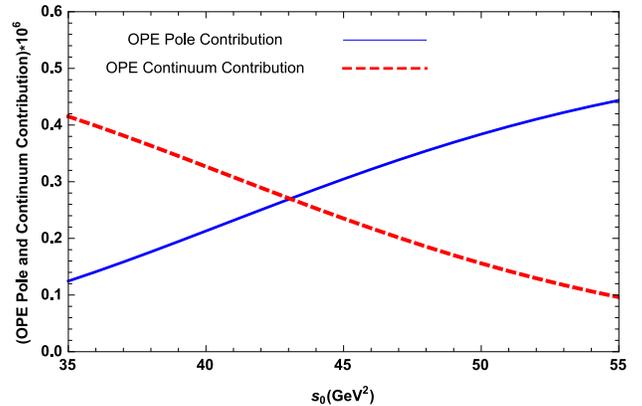}}
\end{picture}
}
\caption{The pole/continuum contribution to the sum rule. By the red (dashed) curve the total higher states and continuum contributions are plotted.}
\label{fig:Pole}
\end{figure}

The working window for the Borel parameter is determined from joint
requirement of the ground state dominance in the sum rule and 
convergence of the relevant operator product expansion. The latter implies  
suppression  of the nonperturbative terms' contributions to the sum rule within
chosen interval for $M^2$.  As a result, for the mass and meson coupling 
calculations we fix the following range for $M^2$
\begin{equation}
3\ \mathrm{GeV}^{2}\leq M^{2}\leq 6\ \mathrm{GeV}^{2}.
\end{equation}%
The choice of the continuum threshold $s_0$ depends on the energy of the first
excited state with the same quantum numbers and content as the particle
under consideration. It can be extracted from comparative analysis of the pole contribution
and higher states-continuum contributions  in the relevant operator
product expansion: The former contribution should overcome
the latter one. Results of the performed numerical computations
are shown in Fig.\ \ref{fig:Pole}.

The analysis performed on the basis of this $\rho^{\mathrm{QCD}}(s)$ allows us to
determine the range of $s_0$ as
\begin{equation}
43\,\,\mathrm{GeV}^{2}\leq s_{0}\leq 45\,\,\mathrm{GeV}^{2}.
\end{equation}%
 
\begin{figure}[tbp]
\centerline{
\begin{picture}(200,170)(0,0)
\put(-10,20){\epsfxsize8.2cm\epsffile{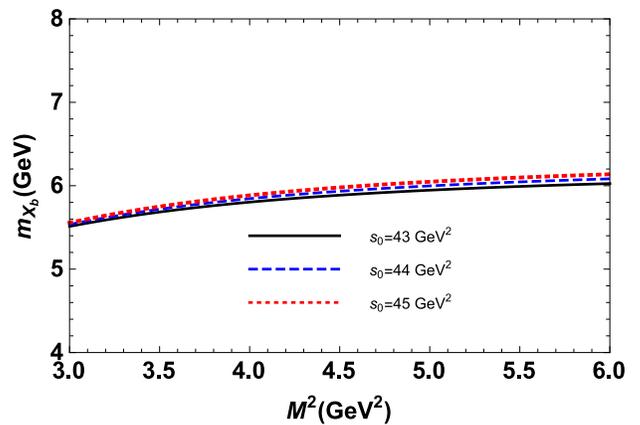}}
\end{picture}}
\caption{The mass $m_{X_b}$ of the axial-vector $X_b$ state vs Borel
parameter $M^2$.}
\label{fig:Mass}
\end{figure}

\begin{figure}[tbp]
\centerline{
\begin{picture}(200,170)(0,0)
\put(-10,20){\epsfxsize8.2cm\epsffile{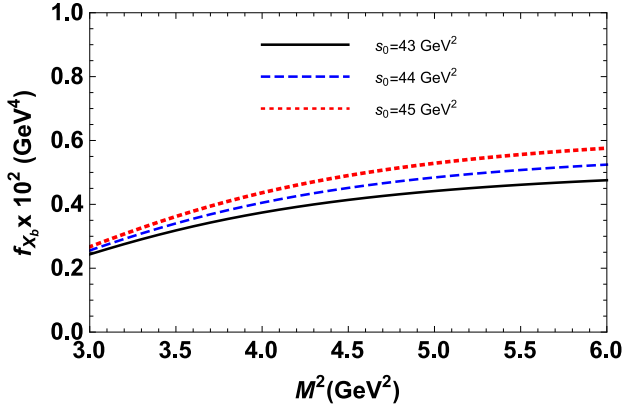}}
\end{picture}}
\caption{The meson coupling $f_{X_b}$ as the function of $M^2$.}
\label{fig:MesCoup}
\end{figure}

With these main parameters at hands we can proceed and carry out computations of 
the mass and meson coupling of the $X_b$ state utilizing corresponding two-point sum rules.  
Our results for $m_{X_{b}}$ and $f_{X_{b}}$ are plotted in
Figs.\ \ref{fig:Mass} and \ref{fig:MesCoup}.
As is seen, in the working regions of $s_0$ and $M^2$ the mass and meson
coupling demonstrate dependencies on these parameters, which are, nevertheless, mild. 
Hence, considering $X_{b}$ as the axial-vector diquark-antidiquark state, 
for its mass we obtain
\begin{equation}
m_{X_{b}}=(5864\pm 158)~\mathrm{MeV},
\end{equation}%
whereas for the meson-current coupling $f_{X_{b}}$ we get
\begin{equation}
f_{X_{b}}=(0.42\pm 0.14)\cdot 10^{-2}~\mathrm{GeV}^{4}.
\end{equation}
To compare our result for the mass of the state $X_b$ with quantum numbers $%
J^P=1^{+}$ with the experimental data of the D0 Collaboration, we first have
to find this value. In accordance with explanations in Ref.\ \cite{D0:2016mwd},
it is defined by the expression
\begin{equation}
m[X(1^{+})]=m[X(0^{+})]+m(B_{s}^{\ast})-m(B_{s}^{0}).
\end{equation}%
This equality, when taking into account $m(B_{s}^{\ast})-m(B_{s}^{0}) \simeq
48.7\ \mathrm{MeV}$ (see, Refs.\ \cite{Agashe:2014kda}), leads to the
experimental value
\begin{equation}
m[X(1^{+})] \simeq 5617\,\mathrm{MeV}.  \label{eq:NewM}
\end{equation}%
In other words, treating $X(5568)$ as an axial-vector particle, we have to consider
Eq.\ (\ref{eq:NewM}) as the data of the D0 Collaboration.
Then it becomes clear that, our prediction for the mass of $X_b$ differs from the
result of the D0 Collaboration given by Eq.\ (\ref{eq:NewM}): Even after
taking into account errors of calculations it overshoots
corresponding experimental result.

\begin{figure}[tbp]
\centerline{
\begin{picture}(200,170)(0,0)
\put(-10,20){\epsfxsize8.2cm\epsffile{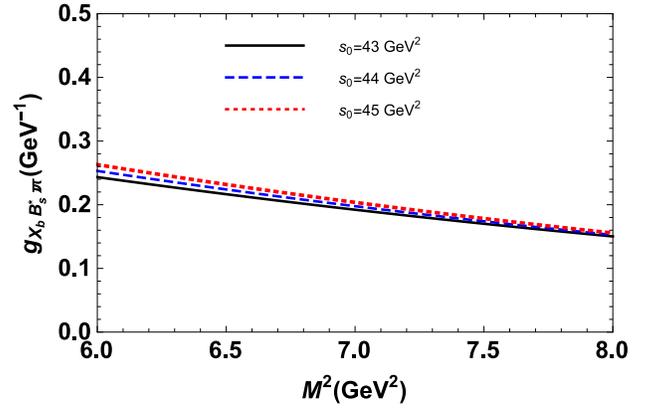}}
\end{picture}}
\caption{Dependence of the strong coupling $g_{X_bB_{s}^{\ast}\protect\pi}$
on the Borel parameter $M^2$.}
\label{fig:VertCoupM}
\end{figure}

\begin{figure}[tbp]
\centerline{
\begin{picture}(200,170)(0,0)
\put(-10,20){\epsfxsize8.2cm\epsffile{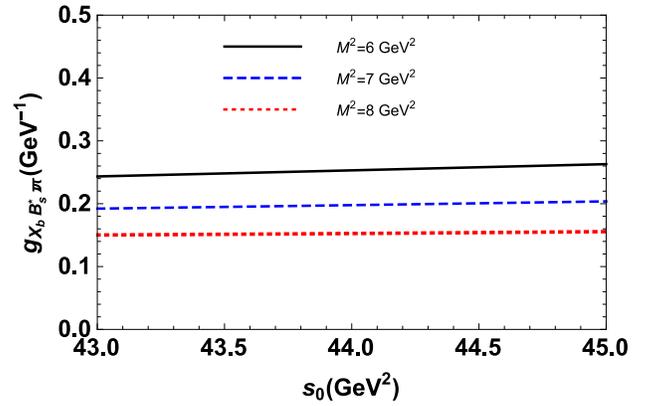}}
\end{picture}}
\caption{The strong coupling $g_{X_bB_{s}^{\ast}\protect\pi}$ as a function
of $s_0$ at some fixed values of $M^2$.}
\label{fig:VertCoupS}
\end{figure}

In evaluation of the strong coupling $g_{X_{b}B_{s}^{\ast}\pi }$ the window
for the Borel parameter shifts towards larger values
\begin{equation}
6\ \mathrm{GeV}^{2}\leq M^{2}\leq 8\ \mathrm{GeV}^{2},
\end{equation}%
whereas the range of $s_0$ remains unchanged. Results of our computations of
the strong coupling $g_{X_{b}B_{s}^{\ast}\pi }$ and its dependence on $M^2$
and $s_0$ are depicted in Figs.\ \ref{fig:VertCoupM} and \ref{fig:VertCoupS}%
, respectively. One can see, that the strong coupling is sensitive to the
choice of the Borel parameter and almost stable under varying of $s_0$.

The coupling $g_{X_{b}B_{s}^{\ast}\pi }$ extracted from the sum rule Eq.\ (%
\ref{eq:SRules1}) reads
\begin{equation}
g_{X_{b}B_{s}^{\ast}\pi }=(0.20\pm 0.05)~\mathrm{GeV}^{-1}.
\end{equation}
Carrying out similar numerical analysis for the coupling  $\widetilde {g}_{X_{b}B_{s}^{\ast}\pi }$
we find:
\begin{equation}
\widetilde{g}_{X_{b}B_{s}^{\ast}\pi}=(0.21\pm 0.06)~\mathrm{GeV}^{-1}.
\end{equation}
The difference between predictions for the couplings
obtained using the different Lorentz structures  is rather small.
We use its mean value $g \Rightarrow (g+\tilde{g})/2$ to evaluate the  width
of the decay $X_{b}\to B_{s}^{\ast}\pi^{+}$ process and get
\begin{equation}
\Gamma(X_{b}\to B_{s}^{\ast}\pi^{+})= (20.2 \pm 6.1) \ \mathrm{MeV},
\end{equation}
whereas the experiment gives $\Gamma_{\mathrm{exp}}[X(5568)\to B_{s}^{\ast}\pi^{+}]=21.9\pm 6.4\mathrm{(stat)}_{-2.5}^{+5.0}\mathrm{(syst)}\,\mathrm{MeV}$.

In other words, the present investigation of the $X(5568)$ resonance and comparison
of the obtained  predictions with data of the D0 Collaboration does not confirm an
axial-vector nature of this diquark-antidiquark state, the mass $m[X_{b}(1^{+})]$
parameter being the decisive argument in making this conclusion.

Within QCD sum rule approach the mass of the $X(5568)$ state as an
axial-vector particle was also investigated in Ref.\ \cite{Chen:2016mqt}.
The result obtained in this paper for the mass of the exotic state $X_b$ reads
\begin{equation}
m[X_{b}(1^{+})]=5.59 \pm 0.15\,\mathrm{GeV}.  \label{eq:wei1}
\end{equation}%
As is seen, there is a discrepancy between our prediction for $%
m[X_{b}(1^{+})]$ and result of Ref.\ \cite{Chen:2016mqt}. The possible source of
the dissonance is the working regions for the parameters $s_0$ and $M^2$ used there,
which were extracted by means of the two-point spectral density
that differs from $\rho ^{\mathrm{QCD}}(s)$ derived in the present work, as it has been noted above.


\section{Concluding remarks}

\label{sec:Conc}
In the current work we have done QCD sum rule analysis of the exotic $X_{b}$
state by considering it as an axial-vector particle built of a diquark and
antidiquark. We have computed the mass $m_{X_{b}}$ and decay width of the
process $X_{b}\to B_{s}^{\ast}\pi^{+}$, and compared our results with
experimental data of the D0 Collaboration, as well as with theoretical
prediction for $m_{X_{b}}$ made in Ref.\ \cite{Chen:2016mqt}. Our result for
the mass of the axial-vector exotic state $X_b$ exceeds the
experimental data, whereas the decay width of the process $%
X_{b}\to B_{s}^{\ast}\pi^{+}$ calculated using two different structures
in the correlation function is compatible the experimental data. Despite 
this last fact, we  make our conclusion relying mainly on the mass calculation: 
If the $X(5568)$ resonance exists and its parameters measured by the D0 
Collaboration are correct, then the present analysis with high
level of confidence excludes, that it is an axial-vector 
diquark-antidiquark state with $J^{P}=1^{+}$.

It is interesting to note, that the same parameters were evaluated  in Refs.\ \cite%
{Agaev:2016mjb,Agaev:2016ijz} using the diquark-antidiquark model for the $%
X_b$ state with quantum numbers $J^{P}=0^{+}$. In these works a satisfactory
agreement with the experimental data of the D0 Collaboration were found.

Further investigations are required to clarify the experimental situation
and theoretical questions surrounding the $X(5568)$ resonance. But possible
outputs of such studies justify efforts paid for their realization, because
the $X(5568)$ state is very intriguing and interesting particle, supposedly
composed of four-quarks of the different flavor, and its investigation may
shed light on many problems of hadron physics.

\section*{ACKNOWLEDGEMENTS}

This work was supported by TUBITAK under the grant no: 115F183.


\appendix*

\section{ The two-point spectral density}

\renewcommand{\theequation}{\Alph{section}.\arabic{equation}} \label{sec:App}

This appendix contains the results obtained for the two-point spectral
density
\begin{equation}
\rho ^{\mathrm{QCD}}(s)=\rho ^{\mathrm{pert.}}(s)+\sum_{k=3}^{8}\rho _{k}(s),
\label{eq:A1}
\end{equation}%
used to calculate from the QCD sum rules the mass and meson coupling of the $%
X_b$ state. In Eqs.\ (\ref{eq:A1}) and (\ref{eq:A2}) by $\rho _{k}(s)$ we
denote the nonperturbative contributions to $\rho ^{\mathrm{QCD}}(s)$. The
explicit expressions for $\rho ^{\mathrm{pert}}(s)$ and $\rho _{k}(s)$ are
presented here as the integrals of the Feynman parameter $z$:
\begin{widetext}
\begin{eqnarray}
&&\rho ^{\mathrm{pert}}(s)=\frac{1}{12288\pi ^{6}}\int\limits_{0}^{a}\frac{%
dzz^{4}}{(z-1)^{3}}\left[ m_{b}^{2}+s(z-1)\right] ^{3}\left[
m_{b}^{2}+5s(z-1)\right], \notag \\
&&\rho _{\mathrm{3}}(s)=\frac{1}{128\pi ^{4}}\int\limits_{0}^{a}\frac{dzz^{2}%
}{(z-1)^{2}}\left[ m_{b}^{2}+s(z-1)\right] \Big \{2\langle \overline{d}%
d\rangle m_{b}\left[ m_{b}^{2}+s(z-1)\right] +m_{s}(z-1)\left( \langle
\overline{s}s\rangle -2\langle \overline{u}u\rangle \right) \left[
m_{b}^{2}+3s(z-1)\right] \Big \},  \notag \\
&&\rho _{\mathrm{4}}(s)=\frac{1}{9216\pi ^{4}}\langle \alpha _{s}\frac{G^{2}%
}{\pi }\rangle \int\limits_{0}^{a}\frac{dzz^{2}}{(z-1)^{3}}\Big \{%
m_{b}^{4}(13z^{2}-21z+9)+2m_{b}^{2}s\left( 23z^{3}-63z^{2}+58z-18\right)
+s^{2}(z-1)^{3}(32z-27)\Big \},  \notag \\
&&\rho _{\mathrm{5}}(s)=\frac{m_{0}^{2}}{192\pi ^{4}}\int\limits_{0}^{a}%
\frac{dzz}{(1-z)}\Big \{3m_{b}\langle \overline{d}d\rangle \left[
m_{b}^{2}+s(z-1)\right] +m_{s}(z-1)(\langle \overline{s}s\rangle -3\langle
\overline{u}u\rangle )\left[ m_{b}^{2}+2s(z-1)\right] \Big \},  \notag \\
&&\rho _{\mathrm{6}}(s)=\frac{1}{61440\pi ^{6}}m_{b}^{2}\langle g_{s}^{3}G^{3}\rangle
\int\limits_{0}^{a}dz\frac{z^{5}}{(1-z)^{3}} +\frac{1}{1296\pi
^{4}}\int\limits_{0}^{a}dz\Bigg \{g_{s}^{2}\langle \overline{d}d\rangle ^{2}z%
\left[ m_{b}^{2}+2s(z-1)\right] +54\pi ^{2}m_{b}m_{s}\langle \overline{d}%
d\rangle (\langle \overline{s}s\rangle -2\langle \overline{u}u\rangle )
\notag \\
&&+z\left[ g_{s}^{2}(\langle \overline{u}u\rangle ^{2}+\langle \overline{s}%
s\rangle ^{2})+108\pi ^{2}\langle \overline{s}s\rangle \langle \overline{u}%
u\rangle \right] \left[ m_{b}^{2}+2s(z-1)\right] \Bigg \}, \notag \\
&&\rho _{\mathrm{7}}(s)=\frac{1}{1152\pi ^{2}}\langle \alpha _{s}\frac{G^{2}%
}{\pi }\rangle \int\limits_{0}^{a}dz\Big \{8m_{b}\langle \overline{d}%
d\rangle +m_{s}\left[ 3z\langle \overline{s}s\rangle +2\langle \overline{u}%
u\rangle (1-4z)\right] \Big \},  \notag \\
&&\rho _{\mathrm{8}}(s)=\frac{11}{73728\pi ^{2}}\langle \alpha _{s}\frac{%
G^{2}}{\pi }\rangle ^{2}\int\limits_{0}^{a}zdz+\frac{1}{24\pi ^{2}}%
m_{0}^{2}\langle \overline{s}s\rangle \langle \overline{u}u\rangle
\int\limits_{0}^{a}(z-1)dz,\label{eq:A2}
\end{eqnarray}%
\end{widetext}
where $a=1-m_{b}^{2}/s$.

\end{document}